\documentclass[prl,twocolumn,showpacs,letter]{revtex4-1}%

\usepackage{amsmath,amsfonts}
\usepackage{graphicx}
\usepackage{epsfig}
\usepackage{dcolumn}
\usepackage{bm}
\usepackage{times}
\usepackage{epstopdf}
\begin{document}

\title{Modelling cytoskeletal traffic: an interplay between passive
diffusion and active transport}
\author{Izaak Neri$^{1, 2}$, Norbert Kern$^{1, 2}$, and Andrea
Parmeggiani$^{1, 2, 3, 4}$} 
\affiliation{$^{1}$Universit\'e Montpellier 2, Laboratoire Charles Coulomb UMR
5221,
F-34095, Montpellier, France\\ 
$^{2}$CNRS, Laboratoire Charles Coulomb UMR 5221, F-34095, Montpellier,
France \\ 
${^3}$  Universit\'e Montpellier 2, Laboratoire DIMNP UMR 5235, F-34095,
Montpellier, France \\ 
${^4}$ CNRS, Laboratoire DIMNP UMR 5235, F-34095, Montpellier, France}
\date{\today}
\begin{abstract}
 We introduce the totally asymmetric exclusion
process with Langmuir kinetics (TASEP-LK) on a network as  a microscopic
model for active motor protein transport on            
the cytoskeleton, immersed in the diffusive cytoplasm.   We discuss how the
interplay between active transport along a network and infinite diffusion in a
bulk reservoir leads to a heterogeneous matter distribution on various scales.
We find three regimes for steady state transport, corresponding to 
the scale of the network, of individual segments or local to sites.  
At low exchange rates  strong
density heterogeneities develop between different segments in the network.  In
this regime one has to consider the topological complexity of the whole network
to
describe transport. In contrast, at moderate exchange rates the
transport through the network decouples, and the physics is
determined by single segments and the local topology.  
At last, for very high exchange rates the homogeneous Langmuir process
dominates the stationary state.  
We introduce effective rate diagrams for the network to identify these different
regimes.
Based on this method we develop an intuitive but generic picture of how the
stationary 
state of excluded volume processes on complex networks can be understood in
terms of
the single-segment phase diagram. 
\end{abstract} 

\maketitle

\paragraph{Introduction}
Statistical physics has been very successful in deducing macroscopic
properties of materials from the interactions  between their microscopic
components. Active matter systems on the other hand, such as active colloids,
animal flocks or cytoskeletal assemblies, are prone to develop out-of
equilibrium
patterns.  These spatial heterogeneities are in fact essential to life processes
 \cite{Ram}.

\begin{figure}[ht!]
  \begin{center}
    \includegraphics[width=.50 \textwidth]{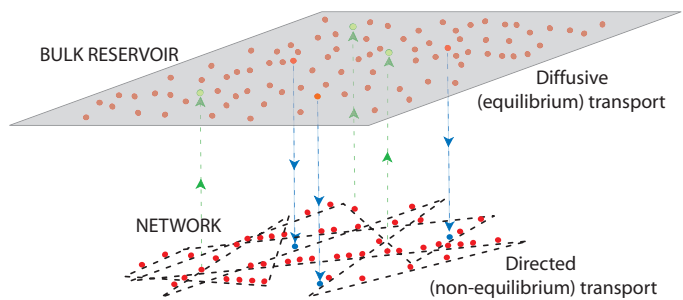}
    \caption{(color online) Statistical physics model of cytoskeletal transport,
      capturing the competition between active transport of particles through a
      network (cytoskeleton) with  diffusion  in a bulk reservoir (cytoplasm). }
    \label{fig:1}
  \end{center}
\end{figure}

Here we would like to initiate a microscopic statistical physics
approach to describe  collective organization of molecular motors in cells.
Motor proteins navigate actively throughout the cell
\cite{Alberts, Ross} along the cytoskeleton, a network of filamentous assemblies
spanning the cytoplasm.   These proteins can
exert forces,  depolymerize filaments and  transport biological cargos along the
cytoskeleton  \cite{Howard}, and thus play an important role in the 
assembly, self-organization and functioning of cells \cite{Fletcher}.  
Single-molecule properties of the motors are well-studied and can now be
measured accurately \cite{Pier}.  But understanding how motors collectively
self-organize remains a very
important step in developing a microscopic vision of intracellular organization.

Macroscopic approaches to study intracellular motor protein transport have been 
developed \cite{Nedelec}, including efforts to introduce microscopic
aspects
\cite{Klumpp,Santen}, 
but a generic microscopic picture of cytoskeletal transport has yet to emerge.
Here we present a tentative approach using a minimalistic model for cytoskeletal
active
transport, which consists of directed motion of particles (motor proteins) along
the
network (cytoskeleton) and diffusion in the bulk (cytoplasm),  see
Fig.~\ref{fig:1}.  We model the directed
motion along the cytoskeleton using the totally
asymmetric simple exclusion process (TASEP) \cite{Derrida} along a
disordered directed network \cite{Neri}.  The
binding and unbinding of particles between the network and
the bulk is represented via Langmuir kinetics (LK) \cite{Langmuir}.  
We consider the particles in the bulk to be infinitely diffusive which, as we
will show, is a relevant limiting case.

TASEP is a fundamental model in non-equilibrium
physics \cite{Mal}, but is  also used in more applied topics, such as
modelling
macromolecules moving through
capillary vessels \cite{Chou1999}, electrons hopping through a quantum-dot chain
\cite{oppen} and vehicular traffic \cite{stad}.  LK  on the other hand  is a
well
known fundamental equilibrium  process in chemical physics
\cite{Langmuir}.

\begin{figure}[ht!]
 \begin{center}
 \includegraphics[width=.50 \textwidth]{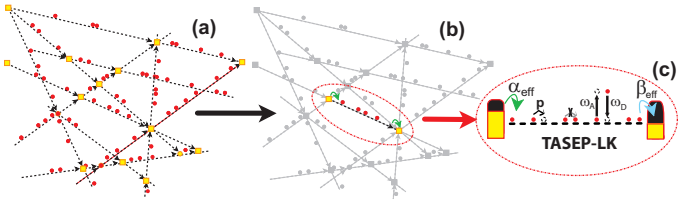}
 \caption{(color online) Sketch of the method to study transport
through complex networks (a), here for TASEP-LK.   Each segment of the
network  (b) is considered to 
connect to two reservoirs with effective rates $\alpha_{\rm eff}$ and
$\beta_{\rm
eff}$ (c),  which are set by the junction densities.
}
\label{fig:graph}
 \end{center}
 \end{figure}

Our model constitutes a generalization of transport through closed networks 
\cite{Neri} to open systems, as they are relevant  for cytoskeletal transport. 
It may also be seen as  generalization to a large scale network of the totally
asymmetric exclusion process with Langmuir kinetics (TASEP-LK) on a single
segment 
\cite{Par03}, which has been shown  to
quantitatively describe in-vitro experiments of motor proteins \cite{Varg}. 
Our study differs from previous work \cite{Klumpp} in that we consider
exclusion interactions and disordered networks, both  
essential points to the physical picture we develop.

The fundamental question we address is how spatial heterogeneities emerge in
such open active systems, here due to the competition between  diffusion  in
a reservoir (which spreads particles homogeneously) and  
active transport  along a network (which generates heterogeneities \cite{Neri}).
  
We will show how this competition is governed by the hopping rate for particles 
on the network, the
exchange rates with the reservoir and the filament length.  One major result of
our work is that 
there exist three regimes of transport through complex networks which are
linked to 
the scale at which spatial density heterogeneities arise: the {\it network},
 the {\it segments} or the {\it sites}. 

 
%

\paragraph{Microscopic model for cytoskeletal transport}
We represent the cytoskeleton as a network of directed segments of $L$ sites
each, 
connected by junction sites, see
Fig.~\ref{fig:graph} (a).  We use random networks to reflect the topological
complexity 
of the cytoskeleton,  a standard approach in modelling networks
\cite{Barrat}. Specifically we use Erd\"os-R\'enyi graphs of mean
connectivity $c$, as in \cite{Neri}.  
Whereas the specific topology  is not important for our qualitative
results,  the fact that networks are {\it irregular}, i.e.~that the number
of incoming $c^{\rm i}_v$ and outgoing  $c^{\rm o}_v$ segments of the junction
$v$ differ, is relevant.  

In each directed segment particles move according to  TASEP-LK
rules \cite{Par03}:   particles hop uni-directionally at rate $p$, subject to
exclusion interactions.  Furthermore, particles obey  binding/unbinding kinetics
with attachment rate $\omega_A$ and detachment rate $\omega_D$ at every
site along the network.  Particles in the reservoir are assumed to diffuse
infinitely fast.

The phase diagram of  TASEP-LK has been determined  \cite{Par03}  
for a single segment connecting a reservoir
with entry rate $\alpha$ to a reservoir with exit rate $\beta$ (see
supplemental materials \cite{Supp1}). 
TASEP-LK is best characterized in terms of the dimensionless  parameters
$\Omega_A=\omega_A\:L/p$ and $\Omega_D=\omega_D\:L/p$, where $\Omega_D$ 
relates the distance an isolated particle typically moves
before detaching to the filament length
\cite{Par03}.   It is convenient to consider the parameters
$\Omega=(\Omega_A+\Omega_D)/2$, characterizing  the total exchange between
reservoir
and segment, and the ratio $K=\Omega_A/\Omega_D$, 
which sets the equilibrium Langmuir density $\rho_\ell = K/(K+1)$.

The $(\alpha, \beta)$-phase diagram of TASEP-LK is fully characterized  in terms
of the parameters
$\Omega$ and $K$  (see \cite{Supp1, Par03} and Fig.~\ref{fig:5}).
In the following we refer to low density (LD), high
density (HD) and maximum current (MC, or M in \cite{Par03}) phases.   All of
these reduce to TASEP phases with a flat
density profile for $\Omega=0$
\cite{Derrida}.
Furthermore  LD and HD zones can coexist on the same segment separated
by a domain wall leading to a coexistence phase (LD-HD).

\paragraph{Mean field method for TASEP-LK on networks}
We analyze TASEP-LK  by extending the mean field arguments for TASEP  presented
in 
\cite{Neri}.  In this approach every
segment $(v, v')$ connecting two vertices $v$ and $v'$ 
is considered to be governed by effective entry and exit rates
$\alpha^{\rm eff}_{(v, v')}$ and $\beta^{\rm eff}_{(v, v')}$,  see
Fig.~\ref{fig:graph} (c).   These rates are  in turn
determined by the average densities $\rho_v$ and $\rho_{v'}$ at the junction
sites
\cite{Em09,Neri}:
$\alpha^{\rm eff}_{v, v'}=p\: \rho_v/c^{\rm o}_v$ and $\beta^{\rm
eff}_{v, v'}=p\: (1-\rho_{v'})$.   
Balancing the currents at the junctions leads to 
the following closed set of equations in $\rho_v$:
\begin{eqnarray}
  \frac{\partial \rho_v}{\partial t} &=&
  \sum_{v'\rightarrow v}J^-\left[\frac{\rho_{v'}}{c^{\rm o}_{v'}},
1-\rho_v\right]
  - \sum_{v'\leftarrow v}J^+
   \left[\frac{\rho_{v}}{c^{\rm o}_{v}}, 1-\rho_{v'}\right] 
  \label{eq:rhoV} 
\end{eqnarray} 
where the sums are over incoming (outgoing) segments, and $J^\pm\left[\alpha/p,
\beta/p\right]$ are the currents entering (leaving) a segment with 
rates $\alpha$ ($\beta$.) 
The expressions for $J^{\pm}$ are readily available
\cite{Supp1,Par03}. Due to the Langmuir process  the current is not constant
along the segment such that $J^-\neq J^+$. 
Solving Eqs.~(\ref{eq:rhoV}) yields the complete stationary state of all
 segments in the network.

The {\it overall} particle density on a network immersed in a reservoir 
is equal to the  Langmuir density $\rho_\ell$, set by the ratio $K$ 
\cite{Neri01}.
In the following we discuss the physical phenomena at fixed $K$ as  the exchange
parameter $\Omega$ smoothly interpolates from purely active transport in
standard
TASEP ($\Omega=0$) to the diffusion dominated limit ($\Omega
\rightarrow \infty)$.  The data shown here has
been
obtained
using $K=1.5$, which is a reasonable value for motor proteins \cite{Reese} and
theoretically not specific in any way: our analysis is general, and only very
high and very low values of
$K$ require an additional discussion \cite{Neri01}.


\begin{figure}[ht!]
  \begin{center}
    \includegraphics[width=.4\textwidth]{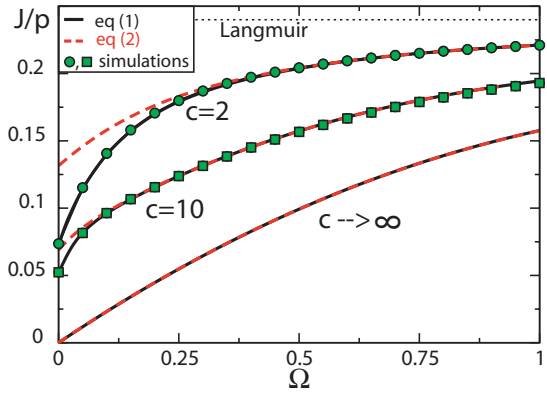}
    \caption{(color online)  
      Average current as a function of the exchange parameter $\Omega$, for
three different (average) connectivities ($c=2$, $c=10$ and $c\rightarrow
\infty$) at $K=1.5$. Agreement between simulation (symbols, for $L=400$) and
mean-field results (solid lines, Eq.~(1)) is excellent. The (red) dashed line is
a simplified mean-field result (Eq.~(2)) while the dotted line denotes the
Langmuir current $J/p = \rho_\ell(1-\rho_\ell)$. 
Results are for a single graph instance of $\mathcal{O}(10^2)$ number of
junctions. 
}
    \label{fig:4}
  \end{center}
\end{figure}

\paragraph{Decoupling due to particle exchange}
In principle the continuity equations (\ref{eq:rhoV})  couples the 
densities   $\rho_{v}$ and $\rho_{v'}$  of those junctions which are linked by a
segment 
$(v,v')$: it is this coupling which makes the transport problem global,
requiring to analyze the whole network simultaneously. 

Here we exploit one  feature of TASEP-LK, which we state by saying that the 
binding-unbinding process can 'decouple' the currents at the segment boundaries.
Indeed,  for the composite LD-HD phase it is known \cite{Par03} that 
the in-current $J^-$ depends on the in-rate  $\alpha^{\rm eff}$ only, whereas
the out-current $J^+$ depends on the out-rate $\beta^{\rm eff}$ only.
Any LD-HD segment therefore lifts one coupling constraint in Eqs. 
(\ref{eq:rhoV}),
since the  in/out currents $J^\pm$ are
determined {\it locally } by the junction densities $\rho_{v}$ and the local
connectivity.

When complete decoupling is achieved, as is expected at high values of $\Omega$,
one can directly deduce the junction densities for an arbitrary network.  
As presented in the supplement \cite{Supp2}, this leads to an exact  solution of
the mean field Eqs.~(\ref{eq:rhoV}). For $K>1$ we have
\begin{eqnarray}
  \rho_v = \left\{ \begin{array}{cc}  c^{\rm o}_v\left(\frac{c^{\rm
i}_v-1
}{c^{\rm o}_vc^{\rm i}_v-1}\right) & \frac{\rho_v}{c^{\rm o}_v}\leq
\frac{1}{K+1};c^{\rm i}_v\neq 1\\ 
 \frac{c^{\rm o}_v}{2} \left(1 -
\sqrt{1-\frac{c^{\rm i}_v}{c^{\rm
o}_v}}\right) &\frac{\rho_v}{c^{\rm o}_v}\leq
\frac{1}{K+1} \\ 
\frac{1}{2}+\sqrt{\frac{1}{4}-\: \frac{c^{\rm o}_v}{c^{\rm
i}_v}
\rho_\ell(1-\rho_\ell)}&\rho_v\geq\frac{c^{\rm o}_v}{K+1}; \rho_v\geq
\frac{1}{2}
\end{array}
\right.\label{eq:simple}
\end{eqnarray}


In Fig.~\ref{fig:4} we compare these analytical predictions based on complete
decoupling to  both a (numerical) full mean-field solution to
Eqs.~(\ref{eq:rhoV}) and to simulations.
For low values of $\Omega$ the fully coupled description Eqs.~(\ref{eq:rhoV}) is
necessary (especially for low mean connectivity $c$).  Surprisingly, 
the decoupled description Eq.~(\ref{eq:simple}) is excellent 
even down to relatively low values of $\Omega$. 
This is an important result, as it shows that single segment TASEP-LK
\cite{Par03} 
suffices to describe transport through any complex network for a wide $\Omega$
range. 
For comparison we have also indicated the average mean-field current in the 
$c\rightarrow \infty$ limit, i.e. the TASEP-LK current which is maintained even
when all
junctions 
are blocked ($\alpha^{\rm eff}=\beta^{\rm eff}=0$).  
This constitutes a lower bound to any current in any network.  


\paragraph{Effective rate plots}
Here we introduce effective rate plots as a way to understand intuitively
the physics of active transport through networks
by allowing to visualize the whole stationary transport state of the network.  
In Figs.~4 we map the effective
rates $(\alpha_s,\beta_s)$  of each segment $s$,  obtained by numerically
solving of the 
full mean field Eqs.~(1), onto the single segment phase
diagram.  Note that the scattering of effective rates is due
to the irregularity of the networks considered.


For sufficiently high $c$ Figs.~\ref{fig:5}  reveal that the effective rates
cluster close to the 
origin, in the LD-HD phase; this explains  why the simplified 
Eqs.~(\ref{eq:simple}) work well  for high $c$ in Fig.\ref{fig:4}. 
When increasing $\Omega$ the TASEP-LK phase diagram
changes. In particular the LD phase reduces in favour of
the LD-HD phase.  Moreover, at high $\Omega$ one notices a
specific alignment of the effective  rates as given by the decoupled
Eqs.~(\ref{eq:simple}). 

In the following section we show how effective rate plots allow to
rationalize the scale at which density heterogeneities appear in the network.

\begin{figure}[ht!]
  \begin{center}
    \includegraphics[width=.5 \textwidth]{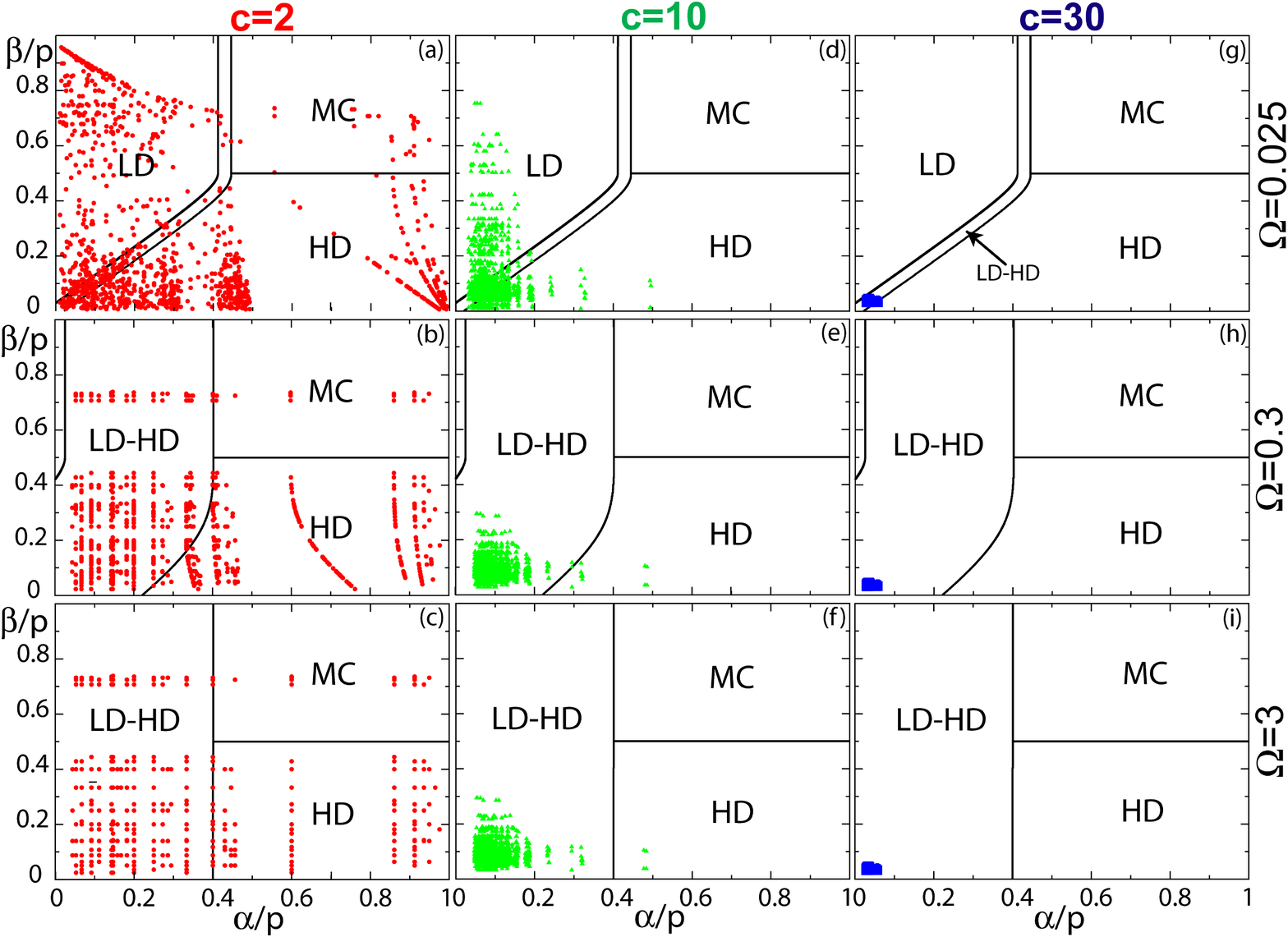}
    \caption{(color online) Effective rate diagrams for 
      irregular graphs of mean connectivity $c$ (at $K=1.5$, for given values of
$\Omega$) .
      Effective rates follow from solving Eqs. (\ref{eq:rhoV}). 
      Single graph instances consisted of $\mathcal{O}(10^2)$ junctions.
    }
    \label{fig:5}
  \end{center}
\end{figure}

\paragraph{Heterogeneities for TASEP-LK on irregular networks}
The parameter $\Omega$ regulates the way particles distribute along the network,
at {\it overall} density $\rho_\ell$, and thus determines how heterogeneities
develop.
We characterize the stationary
state from the effective rate plots by determining the fraction of segments
occupying the corresponding phases, see  Figs.~\ref{fig:6}.  A
complementary point of view is given  in Fig.~\ref{fig:7} with the distribution
$W(\rho_s)$ of the mean segment
densities $\rho_s$ in the network.  From these figures we
conclude that heterogeneities develop throughout the network in three successive
regimes 
(we discuss the case $K>1$ and refer to \cite{Neri01} for $K\leq1$):

\begin{figure}[ht!]
  \begin{center}
    \includegraphics[width=.5 \textwidth]{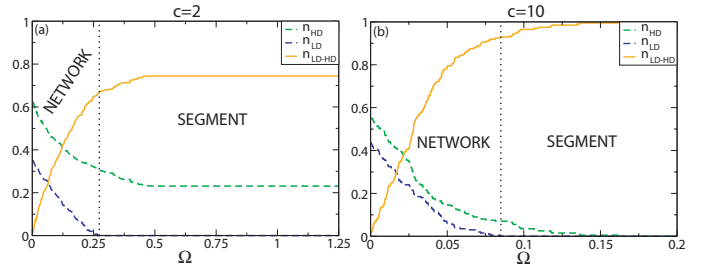}
    \caption{(color online) Fraction of segments in LD, HD and LD-HD
      phases, in the effective rate diagram of mean
      connectivity (a) $c=2$  and (b) $c=10$, for  $K=1.5$. 
      The transition $\Omega_c$ between the network and the segment regime
      (denoted by the vertical dotted line)
      is determined by the condition that $n_{\rm LD}$ vanishes. 
      The graph instances used are those of Fig.  \ref{fig:5}. 
    }
    \label{fig:6}
  \end{center}
\end{figure}

{\it (i)}  the {\it network} regime, for low exchange rates $\Omega$, is
characterized by 
the presence of LD and HD segments. The distribution $W(\rho_s)$ is marked by
the 
LD and HD peaks (whereas LD-HD coexistence segments are distributed evenly over
the 
intermediate density range).
This bimodality implies a strongly heterogeneous density at the network scale.

 {\it (ii)} in the {\it segment} regime, for intermediate exchange rates all 
 LD segments have disappeared in favor of  LD-HD segments.  
 The distribution $W(\rho_s)$ is dominated by the LD-HD peak. 
Although all segments have similar average densities, the presence of domain
walls 
implies strong inhomogeneities on the segment scale.

 {\it (iii)} in the {\it Langmuir} regime, for large exchange rates
$\Omega$,  the Langmuir phase dominates.
All segments have homogeneous densities except for small regions near the
boundaries, and no heterogeneities arise beyond the scale of a few {\it sites}.

\begin{figure}[ht!]
  \begin{center}
    \includegraphics[width=.5 \textwidth]{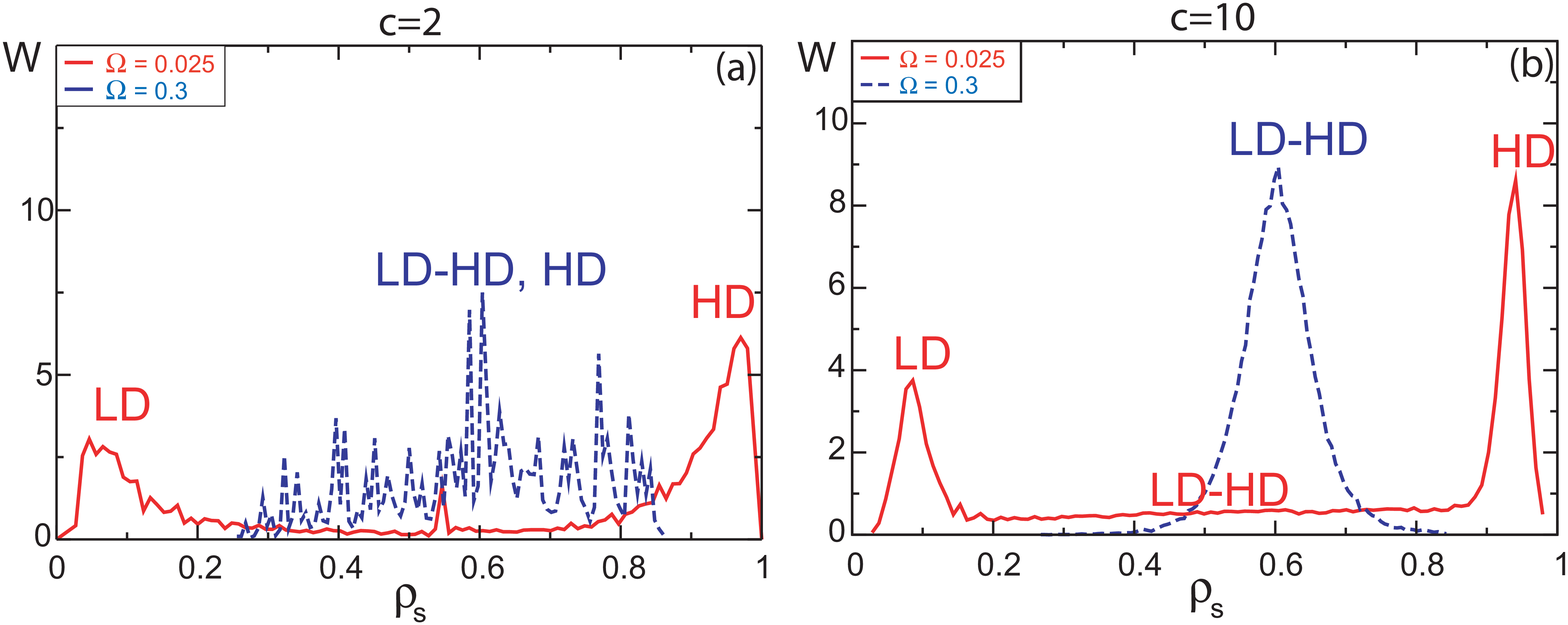}
    \caption{(color online). 
      The MF distribution $W(\rho_s$) of segment
      densities for an irregular graph instance with mean  connectivity (a)
$c=2$ and
      (b) $c=10$, of $\mathcal{O}(10^4)$ junctions, for $K=1.5$. 
      At low values the bimodal distribution of \cite{Neri}
      is identified. When increasing $\Omega$ the center peak gradually grows
while the edge peaks shrink.  Network heterogeneities eventually disappear at
intermediate
      $\Omega=\Omega_c $, leading to a unimodal density distribution.  
    }\label{fig:7}
  \end{center}
\end{figure}

The transition between the network and segment regimes is sharp (identified as
the point $\Omega_c$ where all LD segments disappear, $n_{\rm LD}=0$).  Moreover
this transition has an upper bound $\Omega^\ast = 1/2 +
f(K)\ln[f(K)/(1/2+f(K))]$ (with $f(K) = |K-1|/(2(K+1))$), 
for which the LD phase disappears from the one-dimensional phase diagram
\cite{Neri01}.
In contrast, the  crossover from the segment
regime to the Langmuir regime is progressive. 

\begin{figure}[ht!]
  \begin{center}
    \includegraphics[width=.45 \textwidth]{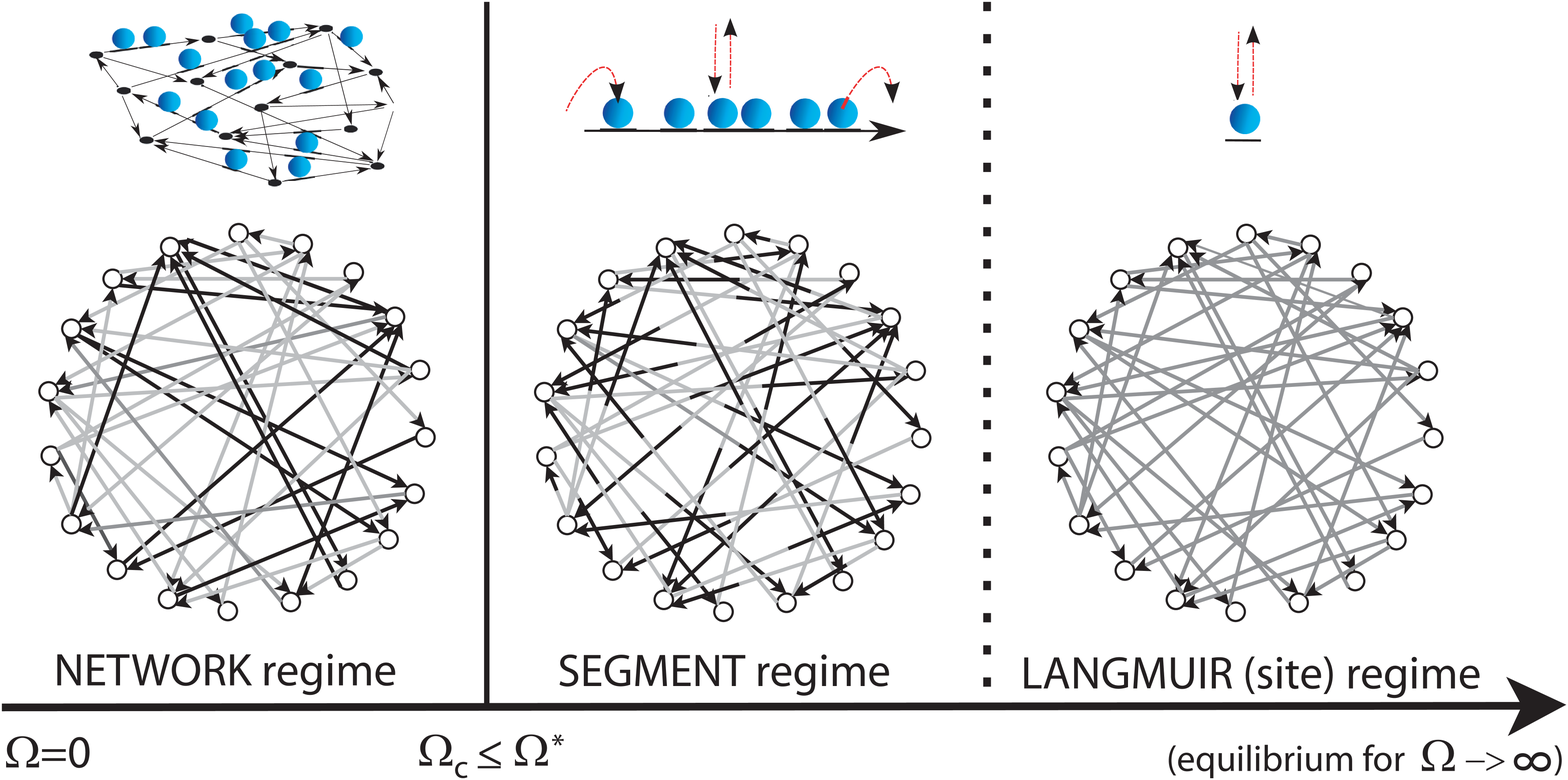}
    \caption{
      Three regimes with
      heterogeneities on a network, segment and site scale, respectively,
      according to the exchange parameter $\Omega$ (see main text for
$\Omega_c$ and $\Omega^\ast$).  Particle densities are coded in grey scale. 
Pictorials on top indicate the scale at which continuity  equations are solved.
    }\label{fig:8}
  \end{center}
\end{figure}

\paragraph{Conclusions}
We have analyzed active transport on a disordered network which is immersed in
an
infinitely diffusive bulk reservoir, as may be considered a simple model for 
cytoskeletal transport by molecular motors. The  dimensionless parameter
$\Omega$   characterizes the  competition between active and passive
transport, and how the associated collective effects lead to  strong spatial
inhomogeneities.

Three regimes arise, according to the scale at which
hetereogeneities appear in the network: a {\it network}, 
a {\it segment} and a {\it Langmuir}  (site-dominated) regime, see Fig.
\ref{fig:8}.
Interestingly, these scales also set the complexity that characterizes
the theoretical
analysis.  
In the network regime TASEP-LK transport is coupled throughout the network,
whereas in the opposite Langmuir regime the physics is essentially determined by
the 
attachment/detachment process. 
In the intermediate segment regime decoupling implies that
the transport characteristics of each segment follow readily from those of  a
single
segment. 

Effective rate plots allow to intuitively understand transport
processes through networks from the single-segment transport characteristics; 
from the scattering of rates over both the LD and HD zones we can directly
deduce the role of strong heterogeneities, see Fig. \ref{fig:5}.  
This approach yields valuable {\it a priori} insight into yet more complex 
excluded volume transport such as TASEP
with extended particles \cite{Zia}, TASEP with multiple species \cite{multiple}
and, as we show in \cite{Neri01}, bidirectional motion \cite{Sandow}:  
as in TASEP-LK, the single-segment phase diagram serves as a basis for 
deducing the behaviour on the network.  
From the effective rate diagram approach it becomes also clear that our results
extend to types of disorder other than topological which are relevant to
biological
systems, e.g.~disorder in the actions of particles at the junctions
\cite{Em09,Raguin}. 
An interesting open question is how finite diffusion \cite{Ebbing}
could be handled within our approach.


Several conclusions may be  relevant for modelling cytoskeletal
transport. 
First, we have shown that strong inhomogeneities in the spatial distribution of
motor proteins form for a wide range of parameters. Even in the  case of
infinitely fast diffusion
considered here they resist the equalizing 
effect of bulk diffusion. Inhomogeneities would therefore be even more relevant
for finite
diffusion. 
Second, the presence of some exchange of motors between the cytoskeleton and the
cytoplasm may in fact simplify a theoretical description, since the
approximation of decoupling Eqs.~(2) yields excellent  results for large enough
$\Omega>\Omega_c$ (see Fig. 3 and 7). 
Third, our analysis hints at a way to regulate the spatial distribution of
motors, and therefore their cargos in the cell, by way of modifying the
exchange parameter
$\Omega$. 
It can be controled in independent ways, via $\omega_D$ or
$\omega_A$ (through 
the bare biochemical rate or through the motor concentration), or via the
length 
dependence in $\Omega$:  regulating the cytoskeleton mesh size, for example by
crosslinker proteins, would allow to control the length scale of
heterogeneities. Values reported in the literature  
\cite{Reese,Varg} show that the values used here ($\Omega, K \sim
\mathcal{O}(1)$)  
is a reasonable order of magnitude, and it is therefore tempting
to speculate that a moderate regulation of $\Omega$  might indeed allow to
provoke a
crossover between the various regimes in living cells.


%
\begin{acknowledgments}
We acknowledge support
from ANR-09-BLAN-0395-02 and from the Scientific Council of the University of
Montpellier 2. We thank C. Leduc for discussing and useful references.  
\end{acknowledgments}

\end{document}